\documentclass[aps,preprint]{revtex4}
\usepackage{amsfonts}
\usepackage{amsmath}
\usepackage{amssymb}
\usepackage{graphicx}

\input{tcilatex}

\begin{document}

\title{\textbf{Quadrupole transitions near interface: general theory and
application to atom inside a planar cavity}}
\author{V.V.Klimov}
\affiliation{P.N.Lebedev Physical Institute, Russian Academy of Sciences, 53 Leninskii
Prospect, 119991 Moscow , Russia}
\email{ vklim@sci.lebedev.ru}
\author{M.Ducloy}
\affiliation{Laboratoire de Physique des Lasers, UMR CNRS 7538 Institut Galilee,
Universite Paris-Nord, Avenue J-B. Clement, F 93430 Villetaneuse, France}

\begin{abstract}
Quadrupole radiation of an atom in an arbitrary environment is investigated
within classical as well as quantum electrodynamical approaches. Analytical
expressions for decay rates are obtained in terms of Green's function of
Maxwell equations. The equivalence of both approaches is shown. General
expressions are applied to analyze the quadrupole decay rate of an atom
placed between two half spaces with arbitrary dielectric constant. It is
shown that in the case when the atom is close to the surface, the total
decay rate is inversely proportional to the fifth power of distance between
an atom and a plane interface.
\end{abstract}

\maketitle

\section{Introduction}

\label{introduction1}

In the recent years the goal of science now has been moving from
consideration of the fundamental properties of atoms to controlling and
changing these properties. It is well known that decay rates of atoms can be
changed in cavities \cite{Purcell}. Many investigations, have been devoted
to a description of the cavity QED effects \cite{Berman}. The main attention
was paid there to the allowed (dipole) transitions. The decay rates of
dipole transitions were investigated in the vicinity of spherical, cylinder,
cone, spheroid, aperture, and more complicated nanobodies \cite{Klimov1},%
\cite{Klimov2}.

However, the influence of environment on the forbidden (quadrupole)
transitions is also of great interest. First of all, it can help one to
study the forbidden transitions. Second, with the help of the forbidden
transitions one can describe the long-living states, which are, in turn,
very important in many applications (quantum computers, quantum
information). Finally, the atoms or molecules with forbidden (quadrupole)
transitions can be used as detectors of field inhomogeneites.

The first experiment dealing with quadrupole transitions near a plane
interface was carried out not along ago. The influence of interface on the
absorption of Cs $6^{2}S_{1/2}-5^{2}D_{5/2}$ transition was studied \cite%
{Tojo1},\cite{Tojo2},\cite{Tojo3}. As for the theoretical works, there were
very few analyses on this topic. In \cite{Chance1},\cite{Chance2}, the
classical calculations of decay rates of quadrupole transitions near the
plane dielectric interface were performed. The quadrupole transitions near
sphere and cylinder were considered within both the classical and QED
approaches, and it was shown that both approaches gave identical results, as
shown in \cite{Klimov3},\cite{Klimov4} . However, there was no exact proof
of equivalence between the classical and QED pictures.

The aim of this paper is to find expressions for the quadrupole decay rates
and to prove their equivalence in an arbitrary environment. In Section \ref%
{ClassicalSection} we derive expressions for total decay rate of a Lorenz
oscillator in arbitrary environment through Green function of Maxwell
equations. In Section \ref{QEDsection} we find the expressions for total
decay rate of an atom in arbitrary environment within the QED approach, and
show that they are the same as those in the classical approach. Then we
apply general results to find expressions for quadrupole decay rates for an
atom placed in a planar cavity (Section \ref{planarcavity4}) This problem is
very important for some experiments on reflection spectroscopy in thin cells 
\cite{Ducloy1}. General expressions for quadrupole decay rates in planar
cavity are investigated for the case of dielectric or metallic walls in
Section \ref{analysis5}.

\section{Classical description of quadrupole radiation in any environment}

\bigskip \label{ClassicalSection}

It is easy to show within classical electrodynamics that the total rate of
work performed by the field can be presented in the form of \cite{Stratton}

\begin{equation}
{\frac{{dE}}{{dt}}}=\int {d^{3}\mathbf{rJ}\left( \mathbf{r}{,t}\right) \cdot 
\mathbf{E}\left( \mathbf{r}{,t}\right) }  \label{eq1}
\end{equation}

\noindent where $J\left( {\mathrm{\mathbf{r}},t} \right)$ and $E\left( {%
\mathrm{\mathbf{r}},t} \right)$ are the density of current and strength of
the electric field, respectively. This power represents a conversion of
electromagnetic field into mechanical or thermal energy. In quasi
monochromatic case we have instead of (\ref{eq1}) the following expression

\begin{equation}
{\frac{{dE}}{{dt}}}={\frac{{1}}{{2}}}Re\int {d^{3}\mathbf{rJ}^{\ast }\left( 
\mathbf{r}{,\omega }\right) \cdot \mathbf{E}\left( \mathbf{r}{,\omega }%
\right) }  \label{eq2}
\end{equation}

\noindent where $\omega $ is the frequency, and * means the complex
conjugation.

The electric field can be expressed through current with the help of
retarded Green function:

\begin{equation}
E_{i}\left( {\mathrm{\mathbf{r}};\omega }\right) ={\frac{{i}}{{\omega }}}%
\int {d^{3}\mathrm{\mathbf{{r}^{\prime }}}}G_{ij}^{R}\left( \mathbf{r}{,%
\mathbf{r}\mathrm{\mathbf{^{\prime }}};\omega }\right) J_{j}\left( {\mathrm{%
\mathbf{{r}^{\prime }}};\omega }\right)  \label{eq3}
\end{equation}

Here and below the lower Latin subscripts denote Cartesian coordinates and
are to be summed over when repeated.

The retarded Green function (\ref{eq3}) is the solution of Helmholtz wave
equation

\begin{equation}
\nabla \times \left( {\nabla \times \overleftrightarrow{G}}^{R}{\left( {%
\mathrm{\mathbf{r}},\mathrm{\mathbf{r}}}^{\prime }{;\omega }\right) }\right)
-\left( {{\frac{{\omega }}{{c}}}}\right) ^{2}\varepsilon \left( {\mathrm{%
\mathbf{r}}}\right) \overleftrightarrow{G^{R}}\left( {\mathrm{\mathbf{r}},%
\mathrm{\mathbf{r}}}^{\prime }{;\omega }\right) =4\pi \left( {{\frac{{\omega 
}}{{c}}}}\right) ^{2}\overleftrightarrow{1}\delta {\mathrm{\mathbf{r}}-%
\mathrm{\mathbf{r}}}^{\prime }  \label{eq4}
\end{equation}

\noindent where $\varepsilon \left( {\mathrm{\mathbf{r}}} \right)$ stands
for dielectric constant of environment, and for simplicity we assume that
the media are nonmagnetic and nondispersive.

Substituting (\ref{eq3}) in (\ref{eq2}) the expression for power can be
presented in the form

\begin{equation}
{\frac{{dE}}{{dt}}}=-{\frac{{1}}{{2\omega }}}Im\int {d^{3}\mathrm{\mathbf{r}}%
\int {d^{3}\mathrm{\mathbf{{r}^{\prime }}}}J_{i}^{\ast }\left( {\mathrm{%
\mathbf{r}},\omega }\right) G_{ij}^{R}\left( \mathbf{r}{,\mathbf{r}\mathrm{%
\mathbf{^{\prime }}};\omega }\right) J_{j}\left( {\mathrm{\mathbf{{r}%
^{\prime }}};\omega }\right) }  \label{eq5}
\end{equation}

To compare classical and quantum calculations it is convenient to consider
stored energy as $E_{0} = \hbar \omega $. As a result the expression for
decay rate will take the following form

\begin{equation}
\gamma _{tot}^{class}={\frac{{1}}{{E_{0}}}}{\frac{{dE}}{{dt}}}=-{\frac{{1}}{{%
2\hbar \omega ^{2}}}}Im\int {d^{3}\mathrm{\mathbf{r}}\int {d^{3}\mathrm{%
\mathbf{r}}}^{\prime }J_{i}^{\ast }\left( {\mathrm{\mathbf{r}},\omega }%
\right) G_{ij}^{R}\left( \mathbf{r}{,\mathbf{r}\mathrm{\mathbf{^{\prime }}}%
;\omega }\right) J_{j}\left( {\mathrm{\mathbf{{r}^{\prime }}};\omega }%
\right) }  \label{eq6}
\end{equation}

For the relative decay rate we will have respectively

\begin{equation}
{\frac{{\gamma _{tot}^{class}}}{{\gamma _{tot,0}^{class}}}}={\frac{{Im\int {%
d^{3}\mathrm{\mathbf{r}}\int {d^{3}\mathrm{\mathbf{r}}}^{\prime }J_{i}^{\ast
}\left( {\mathrm{\mathbf{r}},\omega }\right) G_{ij}^{R}\left( \mathbf{r}{,%
\mathbf{r}\mathrm{\mathbf{^{\prime }}};\omega }\right) J_{j}\left( {\mathrm{%
\mathbf{{r}^{\prime }}};\omega }\right) }}}{{Im\int {d^{3}\mathrm{\mathbf{r}}%
\int {d^{3}\mathrm{\mathbf{r}}}^{\prime }J_{i}^{\ast }\left( {\mathrm{%
\mathbf{r}},\omega }\right) G_{ij}^{R,0}\left( \mathbf{r}{,\mathbf{r}\mathrm{%
\mathbf{^{\prime }}};\omega }\right) J_{j}\left( {\mathrm{\mathbf{{r}%
^{\prime }}};\omega }\right) }}}}  \label{eq7}
\end{equation}

\noindent where $\gamma _{tot,0}^{class} $ and $G_{ij}^{R,0} \left( {\mathrm{%
\mathbf{r}},\mathrm{\mathbf{{r}^{\prime}}};\omega} \right)$ are the total
decay rate and the Green function in uniform (free) space, respectively,

\begin{equation}
G_{ij}^{R,0}\left( {\mathrm{\mathbf{r}},\mathrm{\mathbf{{r}^{\prime }}}%
,\omega }\right) ={\left[ {k^{2}\left( {\delta _{ij}-\mathrm{\mathbf{n}}_{i}%
\mathrm{\mathbf{n}}_{j}}\right) {\frac{{1}}{{{\left\vert {\mathrm{\mathbf{r}}%
-\mathrm{\mathbf{{r}^{\prime }}}}\right\vert }}}}+\left( {3\mathrm{\mathbf{n}%
}_{i}\mathrm{\mathbf{n}}_{j}-\delta _{ij}}\right) \left( {{\frac{{1}}{{{%
\left\vert {\mathrm{\mathbf{r}}-\mathrm{\mathbf{{r}^{\prime }}}}\right\vert }%
^{3}}}}-{\frac{{ik}}{{{\left\vert {\mathrm{\mathbf{r}}-\mathrm{\mathbf{{r}%
^{\prime }}}}\right\vert }^{2}}}}}\right) }\right] }e^{ik{\left\vert {%
\mathrm{\mathbf{r}}-\mathrm{\mathbf{{r}^{\prime }}}}\right\vert }}
\label{eq8}
\end{equation}

In (\ref{eq8}), $\mathrm{\mathbf{n}}={\dfrac{{\mathrm{\mathbf{r}}-\mathrm{%
\mathbf{{r}^{\prime }}}}}{{{\left\vert {\mathrm{\mathbf{r}}-\mathrm{\mathbf{{%
r}^{\prime }}}}\right\vert }}}}$ is the unit vector in the direction from
the atom to the observation point and $k=\omega /c$ is the wave vector of
free space.

In the case of quadrupoles the current and charge densities have the
following form

\begin{equation}
\rho ^{Q}={\frac{{1}}{{3!}}}{\sum\limits_{i,j}{Q_{ij}}}\nabla _{i}\nabla
_{j}\delta \left( \mathbf{r}{-\mathbf{r}\mathrm{\mathbf{^{\prime }}}}\right)
\label{eq9}
\end{equation}

\begin{equation}
j_{i}^{Q}={\frac{{i\omega }}{{3!}}}{\sum\limits_{i,j}{Q_{ij}}}\nabla
_{j}\delta \left( \mathbf{r}{-\mathbf{r}\mathrm{\mathbf{^{\prime }}}}\right)
\label{eq10}
\end{equation}

\noindent where \textbf{\textit{r}} is the radius-vector of the observation
point, \textbf{\textit{\ r}}${\mathrm{\mathbf{^{\prime }}}}$ is the
radius-vector of the quadrupole position, and $Q_{ij}$ is the traceless
quadrupole momentum tensor

\begin{equation}
Q_{ij}=\int {d\mathbf{x}\rho \left( \mathbf{x}\right) }\left( {%
3x_{i}x_{j}-x^{2}\delta _{ij}}\right) .  \label{eq11}
\end{equation}

\begin{figure}
\includegraphics[height=6in]{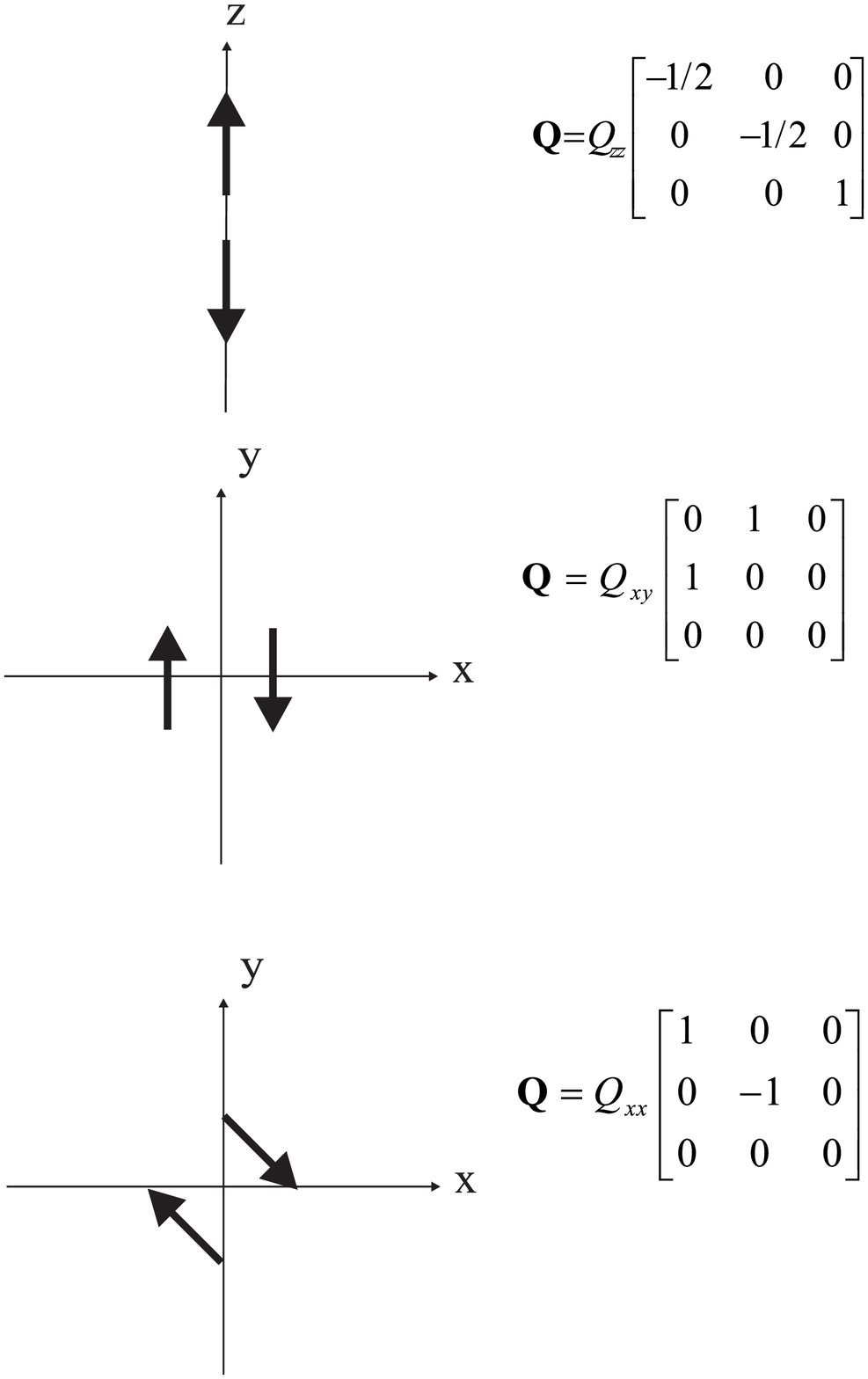}%
\caption{\label{fig1} 	Some kinds of quadrupoles and the corresponding quadrupole momentum tensors.}
\end{figure}

As is known \cite{Stratton}, any quadrupole can be built out of two dipoles
that are equal in amplitude and opposite in orientation. In Fig.\ref{fig1} some
quadrupoles and the related quadrupole momenta are shown.

Now, after a substitution of (\ref{eq10}) into (\ref{eq6}) and (\ref{eq7})
and a partial integration, the final expressions for full and relative decay
rates will take the following form

\begin{equation}
\gamma _{tot}^{class,Q}={\frac{{1}}{{72\hbar }}}\func{Im}{\lim_{\mathrm{%
\mathbf{r}}\rightarrow \mathrm{\mathbf{{r}^{\prime }}}}}Q_{ij}^{\ast
}Q_{kl}\nabla _{j}{\nabla }_{l}^{\prime }G_{ik}^{R}\left( \mathbf{r}{,%
\mathbf{r}\mathrm{\mathbf{^{\prime }}};\omega }\right)  \label{eq12}
\end{equation}

\begin{equation}
{\frac{{\gamma _{tot}^{class,Q}}}{{\gamma _{tot,0}^{class,Q}}}}={\frac{{%
\func{Im}}\underset{\mathrm{\mathbf{r}}\rightarrow \mathrm{\mathbf{{r}%
^{\prime }}}}{{\lim }}{Q_{ij}^{\ast }Q_{kl}\nabla _{j}{\nabla }_{l}^{\prime
}G_{ik}^{R}\left( \mathbf{r}{,\mathbf{r}\mathrm{\mathbf{^{\prime }}};\omega }%
\right) }}{{\func{Im}\underset{\mathrm{\mathbf{r}}\rightarrow \mathrm{%
\mathbf{{r}^{\prime }}}}{{\lim }}Q_{ij}^{\ast }Q_{kl}\nabla _{j}{\nabla }%
_{l}^{\prime }G_{ik}^{R,0}\left( \mathbf{r}{,\mathbf{r}\mathrm{\mathbf{%
^{\prime }}};\omega }\right) }}}  \label{eq13}
\end{equation}

In (\ref{eq12}),(\ref{eq13}) and hereafter, $\nabla ,{\nabla} ^{\prime}$
mean the differentiation over \textbf{r} or \textbf{r$\prime $ ,}
respectively. By calculating the limit in (\ref{eq12}) for free-space Green
function (\ref{eq8}), we obtain that the expression for quadrupole decay
rate in free space gets the following simple form

\begin{equation}
\gamma _{tot,0}^{class,Q}={\frac{{k^{5}}}{{360\hbar }}}{\sum\limits_{ij}{{%
\left\vert {Q_{ij}}\right\vert }}}^{2}  \label{eq14}
\end{equation}

Substituting this expression into (\ref{eq13}) we obtain the following
expression for relative quadrupole decay rate

\begin{equation}
{\frac{{\gamma _{tot}^{class,Q}}}{{\gamma _{tot,0}^{class,Q}}}}=5{\frac{{%
\func{Im}\underset{\mathrm{\mathbf{r}}\rightarrow \mathrm{\mathbf{{r}%
^{\prime }}}}{{\lim }}Q_{ij}^{\ast }Q_{kl}\nabla _{j}{\nabla }_{l}^{\prime
}G_{ik}^{R}\left( \mathbf{r}{,\mathbf{r}\mathrm{\mathbf{^{\prime }}};\omega }%
\right) }}{{k^{5}{\sum\limits_{ij}{{\left\vert {Q_{ij}}\right\vert }}}^{2}}}}
\label{eq15}
\end{equation}

Thus, to calculate quadrupole decay rates it is suffice to determine the
respective derivatives of Green function of Helmholtz wave equation. This
result was quite expected as we know that electric quadrupoles interact with
field gradients.

\section{QED description of quadrupole radiation in any environment: Linear
response theory}

\label{QEDsection}

To calculate the quadrupole decay rate in arbitrary environment we use the
work of ref.\cite{WylieSipe1},\cite{WylieSipe2}, but apply a minimal
coupling Hamiltonian with generalized Coulomb gauge

\begin{equation}
\begin{array}{l}
H_{int}=-{\dfrac{{e}}{{2mc}}}\left( {\mathrm{\mathbf{\hat{p}}}\mathrm{%
\mathbf{\hat{A}}}\left( \mathbf{r}\right) +\mathrm{\mathbf{\hat{A}}}\mathrm{%
\mathbf{\hat{p}}}\left( \mathbf{r}\right) }\right) +{\dfrac{{e^{2}}}{{2mc^{2}%
}}}\mathrm{\mathbf{\hat{A}}}^{2} \\ 
div\left( {\varepsilon \left( \mathbf{r}\right) \mathrm{\mathbf{\hat{A}}}%
\left( \mathbf{r}\right) }\right) =0,\quad \varphi =0%
\end{array}
\label{eq16}
\end{equation}

Here $\mathrm{\mathbf{\hat {p}}}$ is the operator of electron linear
momentum and $\mathrm{\mathbf{\hat {A}}}\left( {\mathrm{\mathbf{r}}} \right)$
is the vector potential at the electron position \textbf{r}. The last term
in (\ref{eq16}) gives no contribution to calculation of decay rates.

Assuming that the matrix element of the electron momentum between initial ${%
\left| {i} \right\rangle} $and final ${\left| {f} \right\rangle} $ states is
zero, that is, ${\left\langle {f} \right|}\mathrm{\mathbf{\hat {p}}}{\left| {%
i} \right\rangle} = 0$ , the Hamiltonian (\ref{eq16}) can be presented in
the form

\begin{equation}
H_{int}=-{\frac{{e}}{{2mc}}}{\frac{{\partial }}{{\partial r_{0,j}}}}%
A_{i}\left( \mathbf{r}^{\prime }\right) \left( {\hat{p}_{i}\left( {r_{j}-r}%
^{\prime }{_{j}}\right) +\left( {r_{j}-r}^{\prime }{_{j}}\right) \hat{p}_{i}}%
\right)  \label{eq17}
\end{equation}

\noindent where $\mathbf{r}^{\prime }$ is the vector of atom position.

In first order , the transition rate from initial atomic state ${\left\vert {%
i}\right\rangle }$to a final state ${\left\vert {f}\right\rangle }$ is given
by Fermi's golden rule \cite{Fermi},

\begin{equation}
\begin{array}{l}
R_{fi}={\dfrac{{2\pi }}{{\hbar }}}\left( {{\dfrac{{e}}{{mc}}}}\right) ^{2}%
\underset{\mathbf{r}\rightarrow \mathbf{r}^{\prime }}{{\lim }}{\dfrac{{%
\partial }}{{\partial r_{j}}}}{\dfrac{{\partial }}{{\partial {r}_{j^{\prime
}}^{\prime }}}}D_{ji}^{fi}D_{{j}^{\prime }{i}^{\prime }}^{if}{%
\sum\limits_{I,F}{p\left( {I}\right) }}{\left\langle {I}\right\vert }%
A_{i}\left( \mathbf{r}\right) {\left\vert {F}\right\rangle }{\left\langle {F}%
\right\vert }A_{{i}^{\prime }}\left( \mathbf{r}{\mathrm{\mathbf{^{\prime }}}}%
\right) {\left\vert {I}\right\rangle }\times \\ 
\delta \left( {E_{F}+E_{f}-E_{I}-E_{i}}\right)%
\end{array}
\label{eq18}
\end{equation}

\noindent where $D_{ji}^{fi}={\left\langle {f}\right\vert }\left( {\left( {%
r_{j}-r}^{\prime }{_{j}}\right) \hat{p}_{i}}\right) {\left\vert {i}%
\right\rangle }$and capital letters denote eigenstates of the rest of the
total system under consideration, neglecting its interaction with the atom
of interest. Such eigenstates might involve, and depend on the coupling
between the radiation field, other atoms, surface excitations, and the like.
For convenience, we refer to those as the \textquotedblleft field
states\textquotedblright . For simplicity we assume here that the field is
in thermal equilibrium at a temperature $T$; $p\left( {I}\right) =\exp
(-\beta E_{I})/{\sum\limits_{k}{\exp (-\beta E_{k})}}$ with $\beta =\left( {%
k_{B}T}\right) ^{-1}$, is the probability that the field is in state I.

Expressing the $\delta $ function of (\ref{eq18}) in the integral form we
find

\begin{equation}
R_{fi}={\frac{{1}}{{\hbar ^{2}}}}\left( {{\frac{{e}}{{mc}}}}\right) ^{2}{%
\lim_{\mathbf{r}\rightarrow \mathbf{r}^{\prime }}}{\frac{{\partial }}{{%
\partial r_{j}}}}{\frac{{\partial }}{{\partial {r}_{j^{\prime }}^{\prime }}}}%
{\int\limits_{-\infty }^{\infty }{dt{\left\langle {A_{i}\left( \mathbf{r}{,t}%
\right) A_{{i}^{\prime }}\left( \mathbf{r}{\mathrm{\mathbf{^{\prime }}},0}%
\right) }\right\rangle }}}D_{ji}^{fi}D_{{j}^{\prime }{i}^{\prime }}^{if}\exp
\left( {i\omega _{0}t}\right)  \label{eq19}
\end{equation}

\noindent where $\omega _{0}=\left( {E_{f}-E_{i}}\right) /\hbar $. In Eq. (%
\ref{eq19}) angular brackets indicate an ensemble average and $\mathrm{A}%
\left( \mathbf{r}{_{0},t}\right) $ is an interaction picture operator,
evolving as if (\ref{eq16}) were not present,

\begin{equation}
\mathrm{\mathbf{A}}\left( \mathbf{r}{,t}\right) =\exp \left( {{\frac{{i}}{{%
\hbar }}}H_{0}t}\right) \mathrm{\mathbf{A}}\left( \mathbf{r}{,t}\right) \exp
\left( {\ -{\frac{{i}}{{\hbar }}}H_{0}t}\right)  \label{eq20}
\end{equation}

In (\ref{eq20}) $H_{0} $ is the Hamiltonian of the whole system without
interaction.

Further, one can rewrite (\ref{eq19}) as Fourier component of the two-point
correlation function, $G_{i{i}^{\prime }}^{A}\left( {\mathrm{\mathbf{r}},%
\mathrm{\mathbf{{r}^{\prime }}};t}\right) ={\left\langle {A_{i}\left( {%
\mathrm{\mathbf{r}},t}\right) A_{{i}^{\prime }}\left( {\mathrm{\mathbf{{r}%
^{\prime }}},0}\right) }\right\rangle }$

\begin{equation}
R_{fi}={\frac{{1}}{{\hbar ^{2}}}}\left( {{\frac{{e}}{{mc}}}}\right) ^{2}{%
\lim_{\mathbf{r}\rightarrow \mathbf{r}^{\prime }}}{\frac{{\partial }}{{%
\partial r_{j}}}}{\frac{{\partial }}{{\partial {r}_{j^{\prime }}^{\prime }}}}%
G_{i{i}^{\prime }}^{A}\left( \mathbf{r}{,}\mathbf{r}^{\prime }{;\omega _{0}}%
\right) D_{ji}^{fi}D_{j^{\prime }i^{\prime }}^{if}  \label{eq21}
\end{equation}

As in our gauge $\mathrm{\mathbf{E}} = - {\frac{{1}}{{c}}}\mathrm{\mathbf{%
\dot {A}}}$ it is possible to show, that (\ref{eq21}) can be presented as

\begin{equation}
R_{fi}={\frac{{1}}{{\hbar ^{2}}}}\left( {{\frac{{e}}{{m\omega _{0}}}}}%
\right) ^{2}{\lim_{\mathbf{r}\rightarrow \mathbf{r}^{\prime }}}{\frac{{%
\partial }}{{\partial r_{j}}}}{\frac{{\partial }}{{\partial {r}_{j^{\prime
}}^{\prime }}}}G_{ii^{\prime }}^{E}\left( \mathbf{r}{,}\mathbf{r}^{\prime }{%
;\omega _{0}}\right) D_{ji}^{fi}D_{j^{\prime }i^{\prime }}^{if}  \label{eq22}
\end{equation}

\noindent where $G_{i{i}^{\prime }}^{E}\left( \mathbf{r}{,\mathbf{r}\mathrm{%
\mathbf{^{\prime }}};t}\right) ={\left\langle {E_{i}\left( \mathbf{r}{,t}%
\right) E_{{i}^{\prime }}\left( \mathbf{r}{\mathrm{\mathbf{^{\prime }}},0}%
\right) }\right\rangle }$is the two-point correlation function of electric
field. It is convenient to express it through the retarded Green function
defined as

\begin{equation}
G_{i{i}^{\prime }}^{R}\left( \mathbf{r}{,\mathbf{r}}^{\prime }{;t}\right) ={%
\frac{{i}}{{\hbar }}}{\left\langle {{\left[ {E_{i}\left( \mathbf{r}{,t}%
\right) E_{{i}^{\prime }}\left( \mathbf{r}^{\prime }{,0}\right) }\right] }}%
\right\rangle }\Theta \left( {t}\right)  \label{eq23}
\end{equation}

In (\ref{eq23}) square brackets mean a commutator and $\Theta \left( {t}
\right)$ is the Heaviside step function.

By applying the fluctuation-dissipation theorem \cite{Davydov} we obtain

\begin{equation}
R_{fi}={\frac{{1}}{{\hbar }}}\left( {{\frac{{e}}{{m\omega _{0}}}}}\right) {%
\lim_{\mathbf{r}\rightarrow \mathbf{r}^{\prime }}}{\frac{{\partial }}{{%
\partial r_{j}}}}{\frac{{\partial }}{{\partial {r}_{j^{\prime }}^{\prime }}}}%
D_{ji}^{fi}D_{j^{\prime }i^{\prime }}^{if}{\frac{{2ImG}_{ii^{\prime }}^{R}{%
\left( \mathbf{r}{,}\mathbf{r}^{\prime }{;\omega _{0}}\right) }}{{{\left[ {%
1-\exp (-\beta \hbar \omega _{0}}\right] }}}}  \label{eq24}
\end{equation}

\noindent where

\begin{equation}
G_{ii^{\prime }}^{R}\left( \mathbf{r}{,\mathbf{r}\mathrm{\mathbf{^{\prime }}}%
;\omega _{0}}\right) ={\int\limits_{-\infty }^{\infty }{dtG_{ii^{\prime
}}^{R}\left( \mathbf{r},{\mathbf{r}}^{\prime }{;t}\right) \exp \left( {%
i\omega _{0}t}\right) }}  \label{eq25}
\end{equation}

\noindent is the Fourier component of retarded Green function of electric
field.

The temperature dependence, which appears in the form of an occupation
number will be important only for $\left( {k_{B}T}\right) \geq \hbar \omega
_{0}$. Since we are interested primarily in the atomic transition energies
of the order of a Rydberg, we can set $T=0$K in this equation. As a result
the quadrupole decay rate will have the following form for $T=0$:

\begin{equation}
R_{fi}={\frac{{2}}{{\hbar }}}\left( {{\frac{{e}}{{m\omega _{0}}}}}\right) {%
\lim_{\mathbf{r}\rightarrow \mathbf{r}^{\prime }}}{\frac{{\partial }}{{%
\partial r_{j}}}}{\frac{{\partial }}{{\partial {r}_{j^{\prime }}^{\prime }}}}%
D_{ji}^{fi}D_{j^{\prime }i^{\prime }}^{if}ImG_{ii^{\prime }}^{R}\left( 
\mathbf{r}{,\mathbf{r}\mathrm{\mathbf{^{\prime }}};\omega _{0}}\right)
\label{eq26}
\end{equation}

As $G_{i{i}^{\prime }}^{R}\left( \mathbf{r}{,\mathbf{r}\mathrm{\mathbf{%
^{\prime }}};\omega }\right) $ describes the response of the system, it is
possible to show that this function is the solution of Maxwell equations 
\cite{Lifshitz}:

\begin{equation}
\nabla \times \left( {\nabla \times \overleftrightarrow{G^{R}}\left( \mathbf{%
r}{,\mathbf{r}}^{\prime }{;\omega }\right) }\right) -\left( {{\frac{{\omega }%
}{{c}}}}\right) ^{2}\varepsilon \left( \mathbf{r}\right) \overleftrightarrow{%
G^{R}}\left( \mathbf{r}{,\mathbf{r}}^{\prime };\omega \right) =4\pi \left( {{%
\frac{{\omega }}{{c}}}}\right) ^{2}\overleftrightarrow{1}\delta \left( 
\mathbf{r}{-\mathbf{r}}^{\prime }\right)  \label{eq27}
\end{equation}

For quadrupole transitions with changing of principal or orbital quantum
number the following identity is true

\begin{equation}  \label{eq28}
{\left\langle {f{\left| {x_{i} {\frac{{\partial}} {{\partial x_{j}}} }}
\right|}i} \right\rangle} = {\frac{{m\omega _{fi}}} {{2\hbar }}}{%
\left\langle {f{\left| {x_{i} x_{j}} \right|}i} \right\rangle}
\end{equation}

Substituting it into eq (\ref{eq24}) and using definition of quadrupole
momentum $Q_{ij}^{fi}=e{\left\langle {(3x_{i}x_{j}-x^{2}\delta _{ij})}%
\right\rangle }_{fi}$we obtain the following expression for decay rates for
arbitrary quadrupole transitions

\begin{equation}
R_{fi}^{Q}={\frac{{1}}{{18\hbar }}}\underset{r\rightarrow {r}^{\prime }}{{%
\lim }}\nabla _{j}\nabla _{l}^{\prime
}Q_{ji}^{fi}Q_{lk}^{if}ImG_{ik}^{R}\left( \mathbf{r}{,\mathbf{r}^{\prime
};\omega }_{0}\right)  \label{eq29}
\end{equation}

\noindent where $G_{jk}^{R}\left( {\mathrm{\mathbf{r}},\mathrm{\mathbf{{r}%
^{\prime }}};\omega }_{0}\right) $ is the retarded Green function of Maxwell
equation (\ref{eq27}).

It is very important to remember that this expression is valid for any
media, including media with losses.

The quadrupole decay rate in free space is described by the same expression
but with free space of the Green function $G_{jk}^{R,0}\left( \mathbf{r},\mathbf{r}^{\prime} ;\omega_{0}\right) $, instead

\begin{equation}
R_{fi}^{Q,0}={\frac{{1}}{{18\hbar }}}\underset{r\rightarrow {r}^{\prime }}{{%
\lim }}\nabla _{j}\nabla _{l}^{\prime
}Q_{ji}^{fi}Q_{lk}^{if}ImG_{ik}^{R,0}\left( \mathbf{r}{,\mathbf{r}^{\prime
};\omega }_{0}\right) ={\frac{{k^{5}}}{{90\hbar }}}{\sum\limits_{ij}{{%
\left\vert {Q_{ij}}\right\vert }}}^{2}  \label{eq30}
\end{equation}

As a result relative decay rates gets the following form

\begin{equation}
{\frac{{R_{fi}^{Q}}}{{R_{fi}^{Q,0}}}}=5{\frac{\underset{r\rightarrow {r}%
^{\prime }}{{\lim }}\nabla _{j}\nabla _{l}^{\prime }{%
Q_{ji}^{fi}Q_{lk}^{if}ImG_{ik}^{R}\left( \mathbf{r}{,\mathbf{r}^{\prime
};\omega }_{0}\right) }}{{k^{5}{\sum\limits_{ij}{{\left\vert {Q_{ij}}%
\right\vert }}}^{2}}}}  \label{eq31}
\end{equation}

Comparing expression (\ref{eq31})with the classical expression (\ref{eq15})
one can see that they are identical. It means that both classical and QED
models are equivalent for description of the total decay rate. Comparison of
(\ref{eq30}) and (\ref{eq14}) reveals the difference by the factor of 4.The
same difference takes place in the case of dipole transitions and is related
to different definitions of dipole and quadrupole momenta in classical and
quantum mechanics.

One should also remember that these equations describe the total decay
rates. To find the radiative decay rates one should use other approaches,
which allow one to take into account the radiation patterns of photons. 
\textit{It can be done, for example, within the classical approach.}

\section{Quadrupole decay rates of an atom placed in a planar cavity}

\label{planarcavity4}

\begin{figure}
\includegraphics[height=5in]{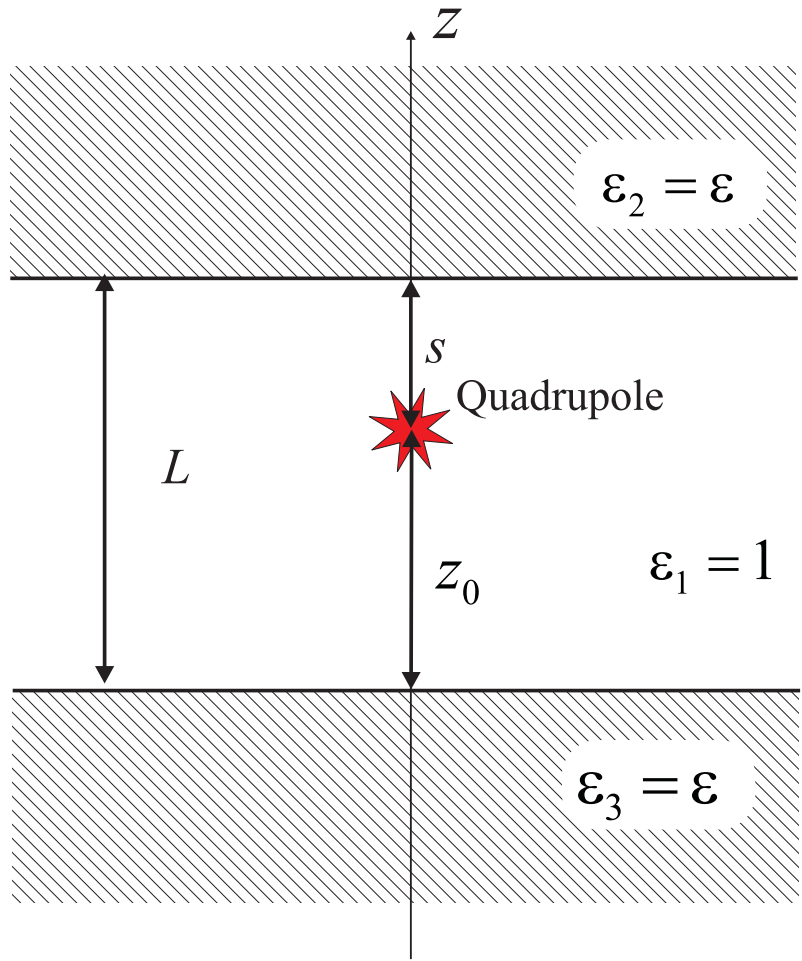}%
\caption{\label{fig2} 	Geometry of the problem of quadrupole radiation of an atom placed in a planar cavity.}

\end{figure}

To calculate the decay rates of quadrupole transition in an atom placed
between two dielectric half-spaces (Fig. \ref{fig2} ) one should find the electric
Green function of Maxwell equation. It is very important to ensure that this
function should satisfy the symmetry condition and the Lorenz reciprocity
relation, which follows from the definition (\ref{eq23}). The approach
suggested in \cite{Tomas} allows one to build such a function. According to 
\cite{Tomas} the Green function in layered media can be presented in the
following form $\left( {z>{z}^{\prime }}\right) $

\begin{equation}
\begin{array}{l}
G\left( {\mathrm{\mathbf{r}},\mathrm{\mathbf{{r}^{\prime }}};\omega }\right)
=\int {{\dfrac{{d^{2}\mathrm{\mathbf{k}}}}{{\left( {2\pi }\right) ^{2}}}}}%
e^{i\mathrm{\mathbf{k}}\left( {\rho -{\rho }^{\prime }}\right) }G\left( {%
\mathrm{\mathbf{k}},z,{z}^{\prime };\omega }\right) \\ 
G\left( {\mathrm{\mathbf{k}},z,{z}^{\prime };\omega }\right) ={\dfrac{{2\pi i%
}}{{\beta _{1}}}\dfrac{{k_{1}^{2}}}{{\varepsilon _{1}}}}e^{i\beta _{1}L}{%
\sum\limits_{q=p,s}{\xi _{q}}}{\dfrac{{E_{q1}^{>}\left( {\mathrm{\mathbf{k}}%
,\omega ,z}\right) E_{q1}^{<}\left( {\ -\mathrm{\mathbf{k}},\omega ,{z}%
^{\prime }}\right) }}{{1-r_{12}^{q}r_{13}^{q}e^{2i\beta _{1}L}}}}%
\end{array}
\label{eq32}
\end{equation}

In (\ref{eq32}), $E_{q1}^{ >} \left( {\mathrm{\mathbf{k}},\omega ,z}
\right),E_{q1}^{ <} \left( {\ - \mathrm{\mathbf{k}},\omega ,{z}^{\prime}}
\right)$ are the mode functions

\begin{equation}
\begin{array}{l}
E_{q1}^{>}\left( {\mathrm{\mathbf{k}},\omega ,z}\right) =\mathrm{\mathbf{%
\hat{e}}}_{q1}^{+}\left( {\mathrm{\mathbf{k}}}\right) e^{i\beta _{1}\left( {%
z-L}\right) }+r_{12}^{q}\mathrm{\mathbf{\hat{e}}}_{q1}^{-}\left( {\mathrm{%
\mathbf{k}}}\right) e^{-i\beta _{1}\left( {z-L}\right) } \\ 
E_{q1}^{<}\left( {\mathrm{\mathbf{k}},\omega ,{z}^{\prime }}\right) =\mathrm{%
\mathbf{\hat{e}}}_{q1}^{-}\left( {\mathrm{\mathbf{k}}}\right) e^{-i\beta
_{1}z}+r_{13}^{q}\mathrm{\mathbf{\hat{e}}}_{q1}^{+}\left( {\mathrm{\mathbf{k}%
}}\right) e^{i\beta _{1}z}%
\end{array}
\label{eq33}
\end{equation}

\noindent and

\begin{equation}
\begin{array}{l}
\mathrm{\mathbf{\hat{e}}}_{p1}^{\pm }\left( {\mathrm{\mathbf{k}}}\right) ={%
\frac{{1}}{{k_{1}}}}\left( {\ \mp \beta _{1}\mathrm{\mathbf{\hat{k}}}+k%
\mathrm{\mathbf{\hat{z}}}}\right) =\mathrm{\mathbf{\hat{e}}}_{p1}^{\mp
}\left( {\ -\mathrm{\mathbf{k}}}\right) \\ 
\mathrm{\mathbf{\hat{e}}}_{s1}^{\pm }\left( {\mathrm{\mathbf{k}}}\right) =%
\mathrm{\mathbf{\hat{k}}}\times \mathrm{\mathbf{\hat{z}}}=-\mathrm{\mathbf{%
\hat{e}}}_{s1}^{\mp }\left( {\ -\mathrm{\mathbf{k}}}\right)%
\end{array}
\label{eq34}
\end{equation}

Here $\beta _{j} = \sqrt {k_{j}^{2} - k^{2}} = \sqrt {\varepsilon _{j}
k_{0}^{2} - k^{2}} (k_{0} = \omega / c)$ is the longitudinal wave vector and 
$r_{12}^{q} ,r_{13}^{q} $ are the conventional Fresnel reflection
coefficients

\begin{equation}  \label{eq35}
r_{ij}^{p} = {\frac{{\varepsilon _{j} \beta _{i} - \varepsilon _{i} \beta
_{j}}} {{\varepsilon _{j} \beta _{i} + \varepsilon _{i} \beta _{j} }}}%
,r_{ij}^{s} = {\frac{{\beta _{i} - \beta _{j}}} {{\beta _{i} + \beta _{j} }}}
\end{equation}

\noindent for p and s polarized waves, and $L$ is the distance between plane
interfaces.

Now by substituting this function into (\ref{eq29}) and integrating it over
angle $\varphi $ in x-y plane ($\hat {k}_{x} = \cos \varphi ,\hat {k}_{y} =
\sin \varphi )$ we obtain the expression of the quadrupole decay rate.

In the case of the z-oriented quadrupole, that is in the case when

\begin{equation}
\mathbf{Q}=Q_{zz}\left[ 
\begin{array}{ccc}
-1/2 & 0 & 0 \\ 
0 & -1/2 & 0 \\ 
0 & 0 & 1%
\end{array}%
\right]  \label{eq36}
\end{equation}

\noindent the decay rate in free space according to (\ref{eq30}) gets the
following form

\begin{equation}
\gamma _{zz}^{0}={\frac{{k_{0}^{5}Q_{zz}^{2}}}{{60}}}  \label{eq37}
\end{equation}

\noindent and the expression for relative decay rate has the following form

\begin{equation}  \label{eq38}
\left( {{\frac{{\gamma}} {{\gamma _{0}}} }} \right)_{zz}^{Q} = {\frac{{15}}{{%
2k_{0}^{5}}} }Re{\int\limits_{0}^{\infty} {k^{3}dk\beta _{1} }} {\frac{{%
\left( {1 - r_{12}^{p} e^{2i\beta _{1} s}} \right)\left( {1 - r_{13}^{p}
e^{2i\beta _{1} z_{0}}} \right)}}{{\left( {1 - r_{12}^{p} r_{13}^{p}
e^{2i\beta _{1} L}} \right)}}}
\end{equation}

In the case of a single interface with ($\left( {L-z_{0}}\right)
=s\rightarrow \infty )$ we have a more simple result \cite{Chance1}

\begin{equation}  \label{eq39}
\left( {{\frac{{\gamma}} {{\gamma _{0}}} }} \right)_{zz}^{Q} = {\frac{{15}}{{%
2k_{0}^{5}}} }Re{\int\limits_{0}^{\infty} {k^{3}dk\beta _{1} }} \left( {1 -
r_{13}^{p} e^{2i\beta _{1} z_{0}}} \right) = 1 - {\frac{{15}}{{2k_{0}^{5}}} }%
Re{\int\limits_{0}^{\infty} {k^{3}dk\beta _{1} }} r_{13}^{p} e^{2i\beta _{1}
z_{0}}
\end{equation}

This coincidence is very interesting because the Green function used for
calculation of decay rate\cite{Chance1} is asymmetric.

In the case of \textit{xy+yx} quadrupole or in the case of \textit{xx-yy}
quadrupole, where

\begin{equation}
\mathbf{Q}=Q_{xy}\left[ 
\begin{array}{ccc}
0 & 1 & 0 \\ 
1 & 0 & 0 \\ 
0 & 0 & 0%
\end{array}%
\right]  \label{eq40}
\end{equation}

\begin{equation}
\mathbf{Q}=Q_{xx}\left[ 
\begin{array}{ccc}
1 & 0 & 0 \\ 
0 & -1 & 0 \\ 
0 & 0 & 0%
\end{array}%
\right]  \label{eq41}
\end{equation}

\noindent the decay rates in free space according to (\ref{eq30}) get the
following form

\begin{equation}
\begin{array}{l}
\gamma _{xy}^{0}={\dfrac{{k_{0}^{5}Q_{xy}^{2}}}{{45}}} \\ 
\gamma _{xx}^{0}={\dfrac{{k_{0}^{5}Q_{xx}^{2}}}{{45}}}%
\end{array}
\label{eq42}
\end{equation}

\noindent and for the relative decay rate we have, respectively,

\begin{equation}
\left( {{\frac{{\gamma }}{{\gamma _{0}}}}}\right) _{xy}^{Q}=\left( {{\frac{{%
\gamma }}{{\gamma _{0}}}}}\right) _{xx}^{Q}={\frac{{5}}{{4k_{0}^{5}}}}Re{%
\int\limits_{0}^{\infty }{{\frac{{k^{3}dk}}{{\beta _{1}}}}}}{\left[ {%
\begin{array}{l}
{\beta _{1}^{2}{\frac{{\left( {1-r_{12}^{p}e^{2i\beta _{1}s}}\right) \left( {%
1-r_{13}^{p}e^{2i\beta _{1}z_{0}}}\right) }}{{\left( {%
1-r_{12}^{p}r_{13}^{p}e^{2i\beta _{1}L}}\right) }}}} \\ 
{\ +k_{1}^{2}{\frac{{\left( {1+r_{12}^{s}e^{2i\beta _{1}s}}\right) \left( {%
1+r_{13}^{s}e^{2i\beta _{1}z_{0}}}\right) }}{{\left( {%
1-r_{12}^{s}r_{13}^{s}e^{2i\beta _{1}L}}\right) }}}}%
\end{array}%
}\right] }  \label{eq43}
\end{equation}

In the case of $s\rightarrow \infty $, that is in the case of single
interface, we have a more simple result \cite{Chance1}

\begin{equation}
\begin{array}{l}
\left( {{\frac{{\gamma }}{{\gamma _{0}}}}}\right) _{zz}^{Q}={\dfrac{{5}}{{%
4k_{0}^{5}}}}Re{\int\limits_{0}^{\infty }{{\dfrac{{k^{3}dk}}{{\beta _{1}}}}}}%
{\left[ {\beta _{1}^{2}\left( {1-r_{13}^{p}e^{2i\beta _{1}z_{0}}}\right)
+k_{1}^{2}\left( {1+r_{13}^{s}e^{2i\beta _{1}z_{0}}}\right) }\right] } \\ 
=1+{\dfrac{{5}}{{4k_{0}^{5}}}}Re{\int\limits_{0}^{\infty }{{\dfrac{{k^{3}dk}%
}{{\beta _{1}}}}}}{\left[ {k_{1}^{2}r_{13}^{s}-\beta _{1}^{2}r_{13}^{p}}%
\right] }e^{2i\beta _{1}z_{0}}%
\end{array}
\label{eq44}
\end{equation}

Finally, in the case of \textit{xz+zx} or \textit{yz+zy} quadrupoles, where

\begin{equation}
\mathbf{Q}=Q_{xz}\left[ 
\begin{array}{ccc}
0 & 0 & 1 \\ 
0 & 0 & 0 \\ 
1 & 0 & 0%
\end{array}%
\right]  \label{eq45}
\end{equation}

\begin{equation}
\mathbf{Q}=Q_{yz}\left[ 
\begin{array}{ccc}
0 & 0 & 0 \\ 
0 & 0 & 1 \\ 
0 & 1 & 0%
\end{array}%
\right]  \label{eq46}
\end{equation}

\noindent the decay rates in free space according to (\ref{eq30}) get the
following form

\begin{equation}
\begin{array}{l}
\gamma _{xz}^{0}={\dfrac{{k_{0}^{5}Q_{xz}^{2}}}{{45}}} \\ 
\gamma _{yz}^{0}={\dfrac{{k_{0}^{5}Q_{yz}^{2}}}{{45}}}%
\end{array}
\label{eq47}
\end{equation}

\noindent and for relative decay rate we respectively have

\begin{equation}
\left( {{\frac{{\gamma }}{{\gamma _{0}}}}}\right) _{xz}^{Q}=\left( {{\frac{{%
\gamma }}{{\gamma _{0}}}}}\right) _{yz}^{Q}={\frac{{5}}{{4k_{0}^{5}}}}Re{%
\int\limits_{0}^{\infty }{{\frac{{kdk}}{{\beta _{1}}}}}}{\left[ {%
\begin{array}{l}
{\left( {\beta _{1}^{2}-k^{2}}\right) ^{2}{\frac{{\left( {%
1+r_{12}^{p}e^{2i\beta _{1}s}}\right) \left( {1+r_{13}^{p}e^{2i\beta
_{1}z_{0}}}\right) }}{{\left( {1-r_{12}^{p}r_{13}^{p}e^{2i\beta _{1}L}}%
\right) }}}} \\ 
{\ +\beta _{1}^{2}k_{1}^{2}{\frac{{\left( {1-r_{12}^{s}e^{2i\beta _{1}s}}%
\right) \left( {1-r_{13}^{s}e^{2i\beta _{1}z_{0}}}\right) }}{{\left( {%
1-r_{12}^{s}r_{13}^{s}e^{2i\beta _{1}L_{1}}}\right) }}}}%
\end{array}%
}\right] }  \label{eq48}
\end{equation}

In the case $s\rightarrow \infty $, that is, in the case of single
interface, we have a more simple result \cite{Chance1}

\begin{equation}
\begin{array}{l}
\left( {{\frac{{\gamma }}{{\gamma _{0}}}}}\right) _{xz}^{Q}=\left( {{\frac{{%
\gamma }}{{\gamma _{0}}}}}\right) _{yz}^{Q}={\dfrac{{5}}{{4k_{0}^{5}}}}Re{%
\int\limits_{0}^{\infty }{{\dfrac{{kdk}}{{\beta _{1}}}}}}{\left[ {\left( {%
\beta _{1}^{2}-k^{2}}\right) ^{2}\left( {1+r_{13}^{p}e^{2i\beta _{1}z_{0}}}%
\right) +\beta _{1}^{2}k_{1}^{2}\left( {1-r_{13}^{s}e^{2i\beta _{1}z_{0}}}%
\right) }\right] }= \\ 
1+{\dfrac{{5}}{{4k_{0}^{5}}}}Re{\int\limits_{0}^{\infty }{{\dfrac{{kdk}}{{%
\beta _{1}}}}}}{\left[ {\left( {\beta _{1}^{2}-k^{2}}\right)
^{2}r_{13}^{p}-r_{13}^{s}\beta _{1}^{2}k_{1}^{2}}\right] }e^{2i\beta
_{1}z_{0}}%
\end{array}
\label{eq49}
\end{equation}

As mentioned above, these results describe the total decay rates, i.e.
radiative and nonradiative. Generally, it is difficult to separate these
contributions. However, one can assume that this separation can be made on
the basis of the classical energy flux method. It should be noted that a
purely radiation channel may exist, in this geometry, in an ideal case of
matter without losses. At negligibly small losses, the radiation energy
would not go to infinity. This is the difference between the geometry under
consideration and an open geometry, at which the radiation might go to
infinity throughout a free space.

\section{Analysis of results and illustrations}

\label{analysis5}

\bigskip

The expressions that had been obtained in the previous section are rather
complicated and their calculation is an independent problem, in a general
case. The complexity is due to the fact that the integrands are the complex
functions with a set of the singular points, which might be both the
branching points, and the poles. These peculiarities are connected with
physical properties of the problem. In any case, in the integrand there are
the branching points at $k = \pm \sqrt {\varepsilon} k_{0} $. If the mode
wave propagation is formed in a cavity (metallic mirrors) then the poles
appear in the integrands. So, in different physical situations, the
calculations are to be performed with account of these factors.

\bigskip

\subsection{\textbf{\textit{Atom between perfect metallic mirrors}}}

\bigskip

In the case of the well conducting metallic mirrors, the expressions (\ref%
{eq38}), (\ref{eq43}), and (\ref{eq48}), in which the reflection
coefficients are substituted by their analogs for the case of an ideal
conductivity,

\begin{equation}
r^{p}=1,r^{s}=-1  \label{eq49a}
\end{equation}

\noindent will be good approximations for the rates

The expressions for the decay rates of quadrupole states may be reduced to
the form

\begin{equation}  \label{eq50}
\left( {{\frac{{\gamma}} {{\gamma _{0}}} }} \right)_{zz}^{Q} = {\frac{{15}}{{%
k_{0}^{5}}} }Im{\int\limits_{0}^{\infty} {k^{3}dk\beta _{1}} }{\frac{{\sin
\left( {\beta _{1} s} \right)\sin \left( {\beta _{1} z_{0}} \right)}}{{\sin
\left( {\beta _{1} L} \right)}}}
\end{equation}

\begin{equation}  \label{eq51}
\left( {{\frac{{\gamma}} {{\gamma _{0}}} }} \right)_{xy}^{Q} = \left( {{%
\frac{{\gamma}} {{\gamma _{0}}} }} \right)_{xx}^{Q} = {\frac{{5}}{{2k_{0}^{5}%
}} }Im{\int\limits_{0}^{\infty} {{\frac{{k^{3}dk}}{{\beta _{1}}} }}} {\left[ 
{\beta _{1}^{2} + k_{1}^{2}} \right]}{\frac{{\sin \left( {\beta _{1} s}
\right)\sin \left( {\beta _{1} z_{0}} \right)}}{{\sin \left( {\beta _{1} L}
\right)}}}
\end{equation}

\begin{equation}  \label{eq52}
\left( {{\frac{{\gamma}} {{\gamma _{0}}} }} \right)_{xz}^{Q} = \left( {{%
\frac{{\gamma}} {{\gamma _{0}}} }} \right)_{yz}^{Q} = - {\frac{{5}}{{%
2k_{0}^{5}}} }Im{\int\limits_{0}^{\infty} {{\frac{{kdk}}{{\beta _{1}}} }}} {%
\left[ {\left( {\beta _{1}^{2} - k^{2}} \right)^{2} + \beta _{1}^{2}
k_{1}^{2}} \right]}{\frac{{\cos \left( {\beta _{1} s} \right)\cos \left( {%
\beta _{1} z_{0}} \right)}}{{\sin \left( {\beta _{1} L} \right)}}}
\end{equation}

\begin{figure}
\includegraphics[height=5in]{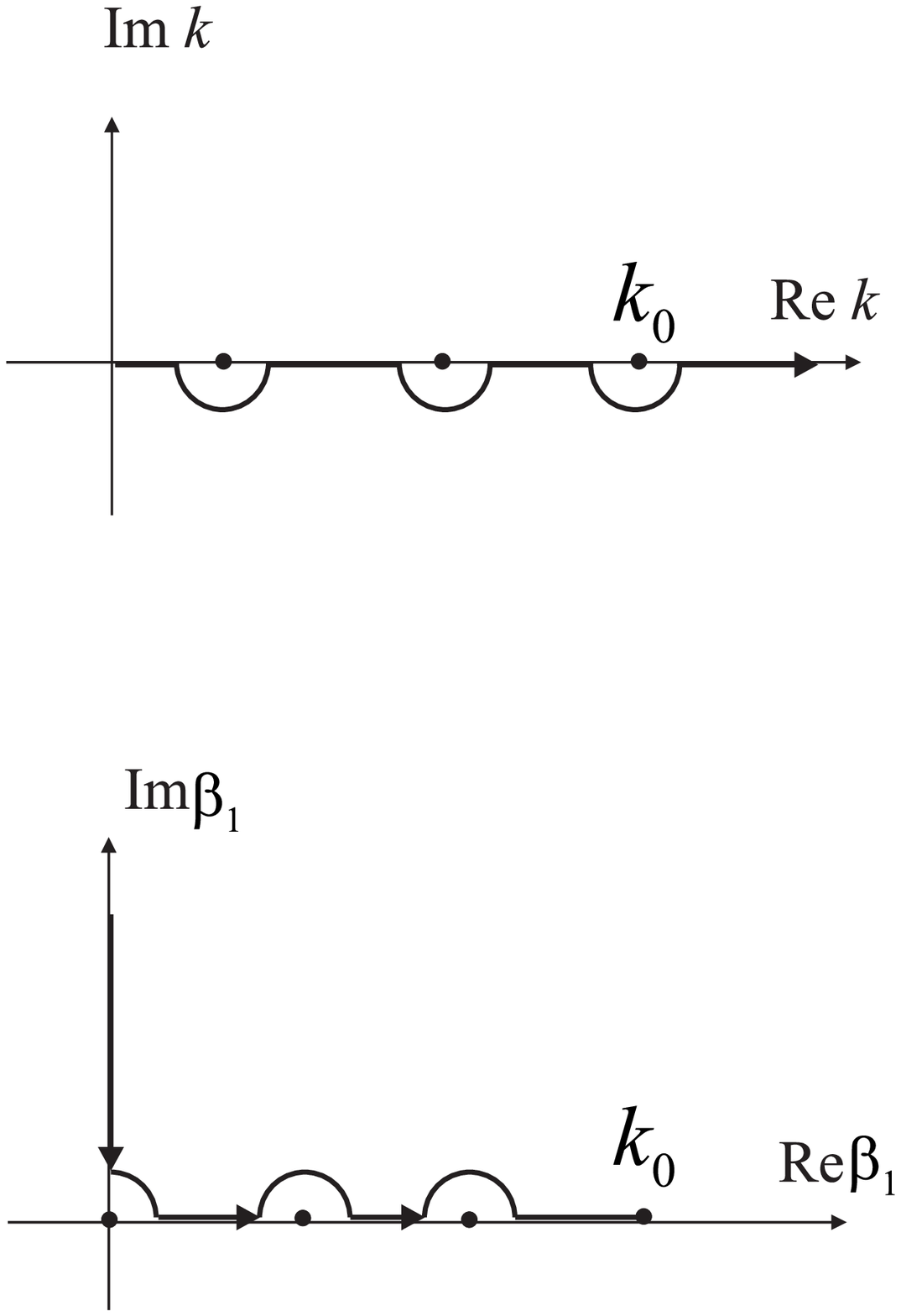}%
\caption{\label{fig3} Contours of integration in the case of ideally conducting mirrors.}

\end{figure}

Because the ideal conductivity is the limiting case for a real metal, where
the poles must lie above the horizontal axis of integration, the integration
circuit of the ideal conductivity must envelope the poles from below, as
shown in Fig.\ref{fig3}.

To calculate the integrals (\ref{eq50})-(\ref{eq52}) it is convenient to use
the variable $\beta _{1} = \sqrt {k_{0}^{2} - k^{2}} $,

\begin{equation}  \label{eq53}
\left( {{\frac{{\gamma}} {{\gamma _{0}}} }} \right)_{zz}^{Q} = {\frac{{15}}{{%
k_{0}^{5}}} }Im{\int\limits_{i\infty} ^{k_{0}} {\left( {k_{0}^{2} - \beta
_{1}^{2}} \right)d\beta _{1} \beta _{1}^{2}} }{\frac{{\sin \left( {\beta
_{1} s} \right)\sin \left( {\beta _{1} z_{0}} \right)}}{{\sin \left( {\beta
_{1} L} \right)}}}
\end{equation}

\begin{equation}  \label{eq54}
\left( {{\frac{{\gamma}} {{\gamma _{0}}} }} \right)_{xy}^{Q} = \left( {{%
\frac{{\gamma}} {{\gamma _{0}}} }} \right)_{xx}^{Q} = {\frac{{5}}{{2k_{0}^{5}%
}} }Im{\int\limits_{i\infty} ^{k_{0}} {\left( {k_{0}^{4} - \beta _{1}^{4}}
\right)d\beta _{1}}} {\frac{{\sin \left( {\beta _{1} s} \right)\sin \left( {%
\beta _{1} z_{0}} \right)}}{{\sin \left( {\beta _{1} L} \right)}}}
\end{equation}

\begin{equation}  \label{eq55}
\left( {{\frac{{\gamma}} {{\gamma _{0}}} }} \right)_{xz}^{Q} = \left( {{%
\frac{{\gamma}} {{\gamma _{0}}} }} \right)_{yz}^{Q} = - {\frac{{5}}{{%
2k_{0}^{5}}} }Im{\int\limits_{i\infty} ^{k_{0}} {d\beta _{1}} }{\left[ {%
\left( {2\beta _{1}^{2} - k_{0}^{2}} \right)^{2} + \beta _{1}^{2} k_{0}^{2}} %
\right]}{\frac{{\cos \left( {\beta _{1} s} \right)\cos \left( {\beta _{1}
z_{0}} \right)}}{{\sin \left( {\beta _{1} L} \right)}}}
\end{equation}

\noindent where the path of integration is shown in Fig.\ref{fig3} . By calculating
the integrals (\ref{eq53})-(\ref{eq55}) with the residue theorem one can
obtain the following results

\begin{equation}  \label{eq56}
\left( {{\frac{{\gamma}} {{\gamma _{0}}} }} \right)_{zz}^{Q} = {\frac{{15\pi 
}}{{\tilde {L}}}}{\sum\limits_{n = 1}^{n_{\max}} {\left( {{\frac{{\pi n}}{{%
\tilde {L}}}}} \right)^{2}\left( {1 - \left( {{\frac{{\pi n}}{{\tilde {L}}}}}
\right)^{2}} \right)}} \sin ^{2}\left( {{\frac{{\pi nz_{0}}} {{L}}}} \right)
\end{equation}

\begin{equation}  \label{eq57}
\left( {{\frac{{\gamma}} {{\gamma _{0}}} }} \right)_{xy}^{Q} = \left( {{%
\frac{{\gamma}} {{\gamma _{0}}} }} \right)_{xx}^{Q} = {\frac{{5\pi }}{{2%
\tilde {L}}}}{\sum\limits_{n = 1}^{n_{\max}} {\left( {1 - \left( {{\frac{{%
\pi n}}{{\tilde {L}}}}} \right)^{4}} \right)}} \sin ^{2}\left( {{\frac{{\pi
nz_{0}}} {{L}}}} \right)
\end{equation}

\begin{equation}  \label{eq58}
\left( {{\frac{{\gamma}} {{\gamma _{0}}} }} \right)_{xz}^{Q} = \left( {{%
\frac{{\gamma}} {{\gamma _{0}}} }} \right)_{yz}^{Q} = {\frac{{5\pi }}{{2%
\tilde {L}}}}{\left[ {{\frac{{1}}{{2}}} + {\sum\limits_{n = 1}^{n_{\max }} {{%
\left\{ {\left( {1 - 2\left( {{\frac{{\pi n}}{{\tilde {L}}}}} \right)^{2}}
\right)^{2} + \left( {{\frac{{\pi n}}{{\tilde {L}}}}} \right)^{2}} \right\}}}%
} \cos ^{2}\left( {{\frac{{\pi nz_{0}}} {{L}}}} \right)} \right]}
\end{equation}

where $n_{\max }=\left[ \widetilde{L}/\pi \right] $ is integral part of $%
\widetilde{L}/\pi $, and $\widetilde{L}=k_{0}L$

\begin{figure}
\includegraphics[height=4in]{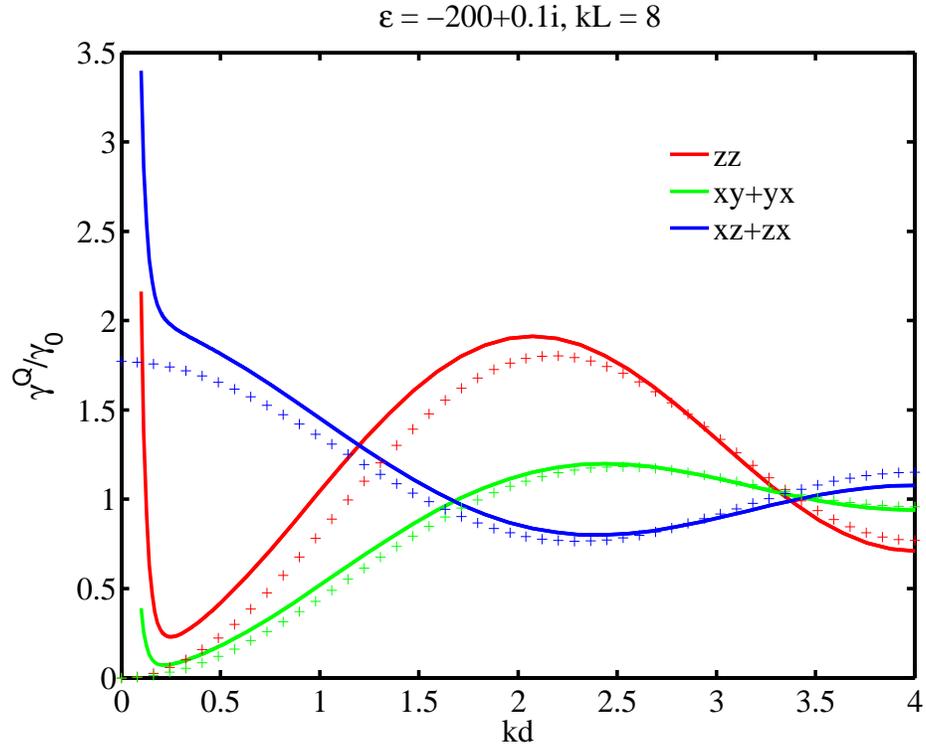}%
\caption{\label{fig4} The quadrupole decay rates of different quadrupoles
 versus their position in the case of a hypothetic material with  $\epsilon=-200+0.1i$, 
    and in the case of an ideal conductivity (dotted line).}
\end{figure}

Figure \ref{fig4}  illustrates the quadrupole decay rates in a resonator formed by a
hypothetic metal with $\varepsilon = - $200+0.01$i$ in respect to the
position and orientation of a quadrupole. As seen from the Figure, the
asymptotic expressions (\ref{eq56})-(\ref{eq58}) approximate well the exact
expressions (\ref{eq38}),(\ref{eq43})(\ref{eq48}), excluding the region that
is in a close proximity to the metal surface. However, in the vicinity of
the surface, the nonradiative losses connected with imaginary part of the
dielectric constant are of the main importance. These losses are the reason
of a fast increase in the total losses (see Eqs.(\ref{eq63})-(\ref{eq65})).
In the case of an ideal conductor, the losses are absent, and there is a
difference between the decay rates of a hypothetic metal and an ideal
conductor.

\begin{figure}
\includegraphics[height=4in]{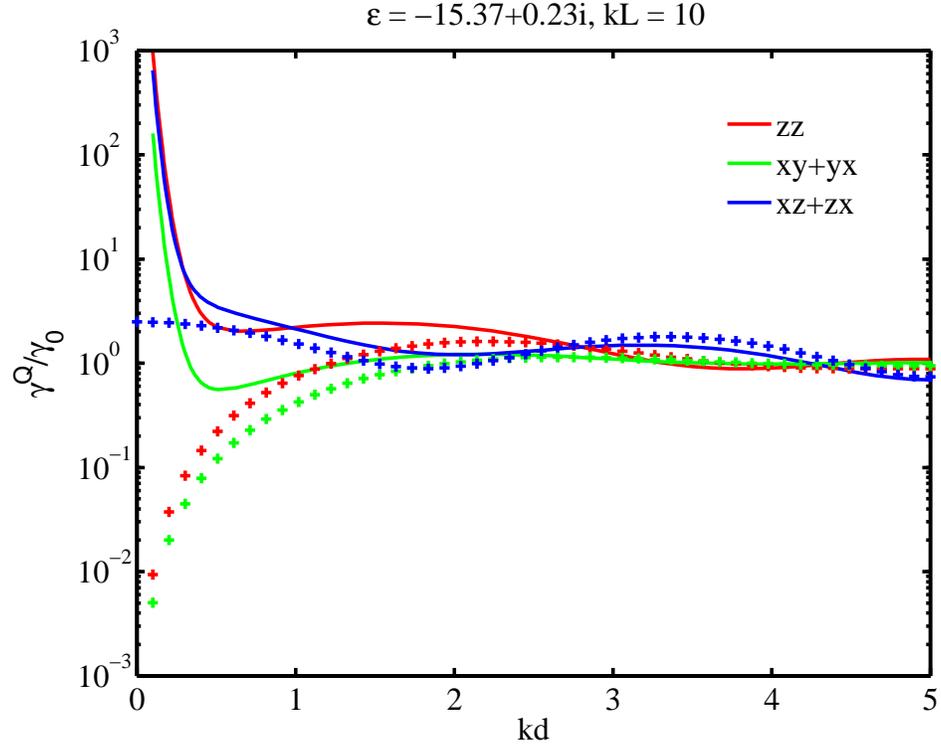}%
\caption{\label{fig5} The quadrupole decay rates of different quadrupoles
 versus their position between two thick silver 
($Ag:\epsilon= –15.37 + 0.231i$ ,$\lambda=632.8 nm$ \cite{Hass}) mirrors 
(dotted lines correspond to the case of the ideally conducting walls).}
\end{figure}

In the case of real metals, that difference might be still more profound
because the imaginary part of the permittivity is not negligibly small as
compared to the real part. Figure \ref{fig5}  illustrates the decay rates for a
micro-resonator with silver mirrors. From the Figure one can see that the
rate of spontaneous decays in the real resonator differs from the decay rate
in the cavity with ideal walls substantially.

\subsection{Atom between dielectric mirrors}

\begin{figure}
\includegraphics[height=4in]{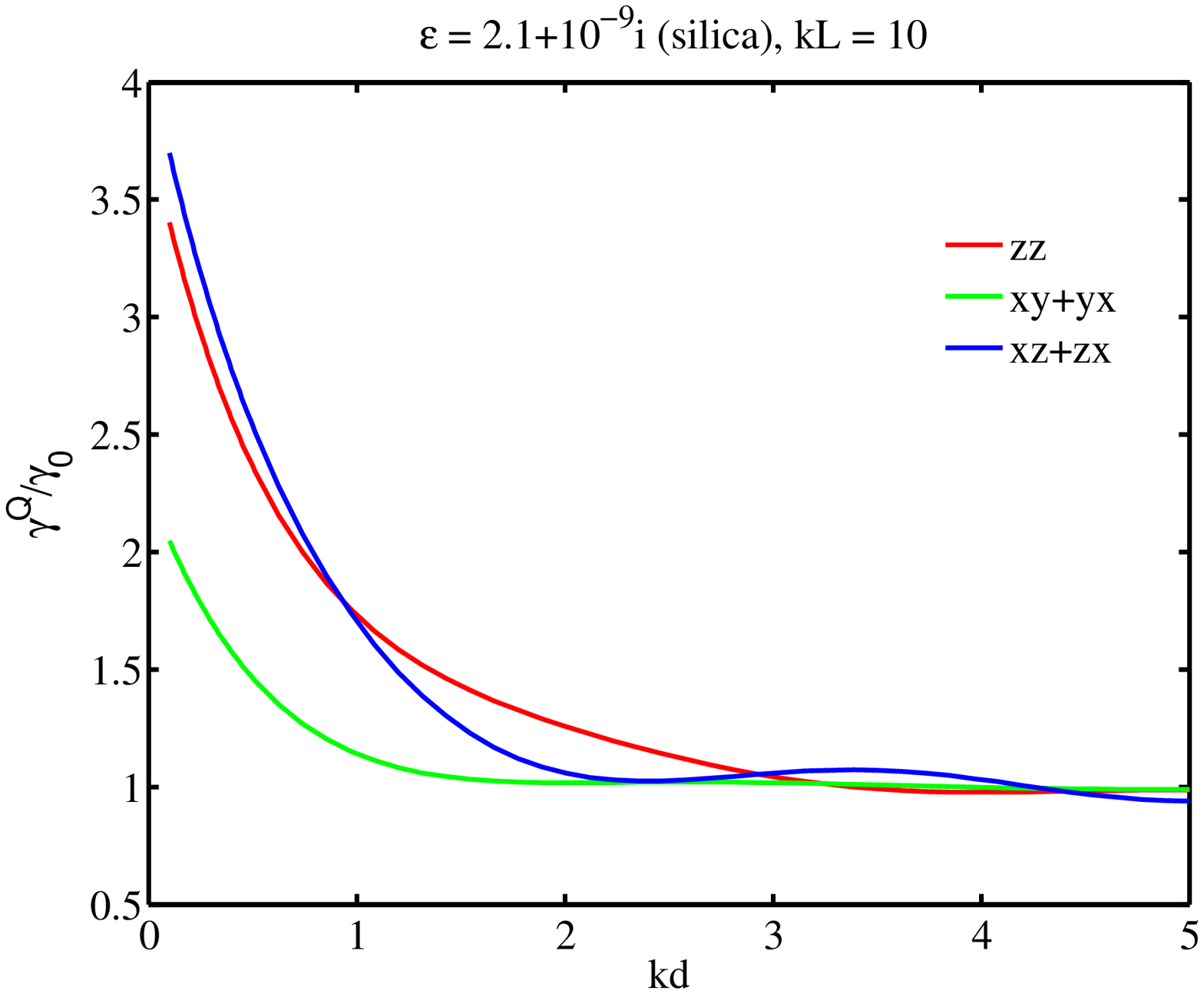}%
\caption{\label{fig6} 	The spontaneous decay rates of different 
quadrupoles versus their position between two quartz half-spaces with 
$\epsilon=2.1+0.000000001i $(silica) in the case of a micro-resonator ($kL=10$). }
\end{figure}

\begin{figure}
\includegraphics[height=4in]{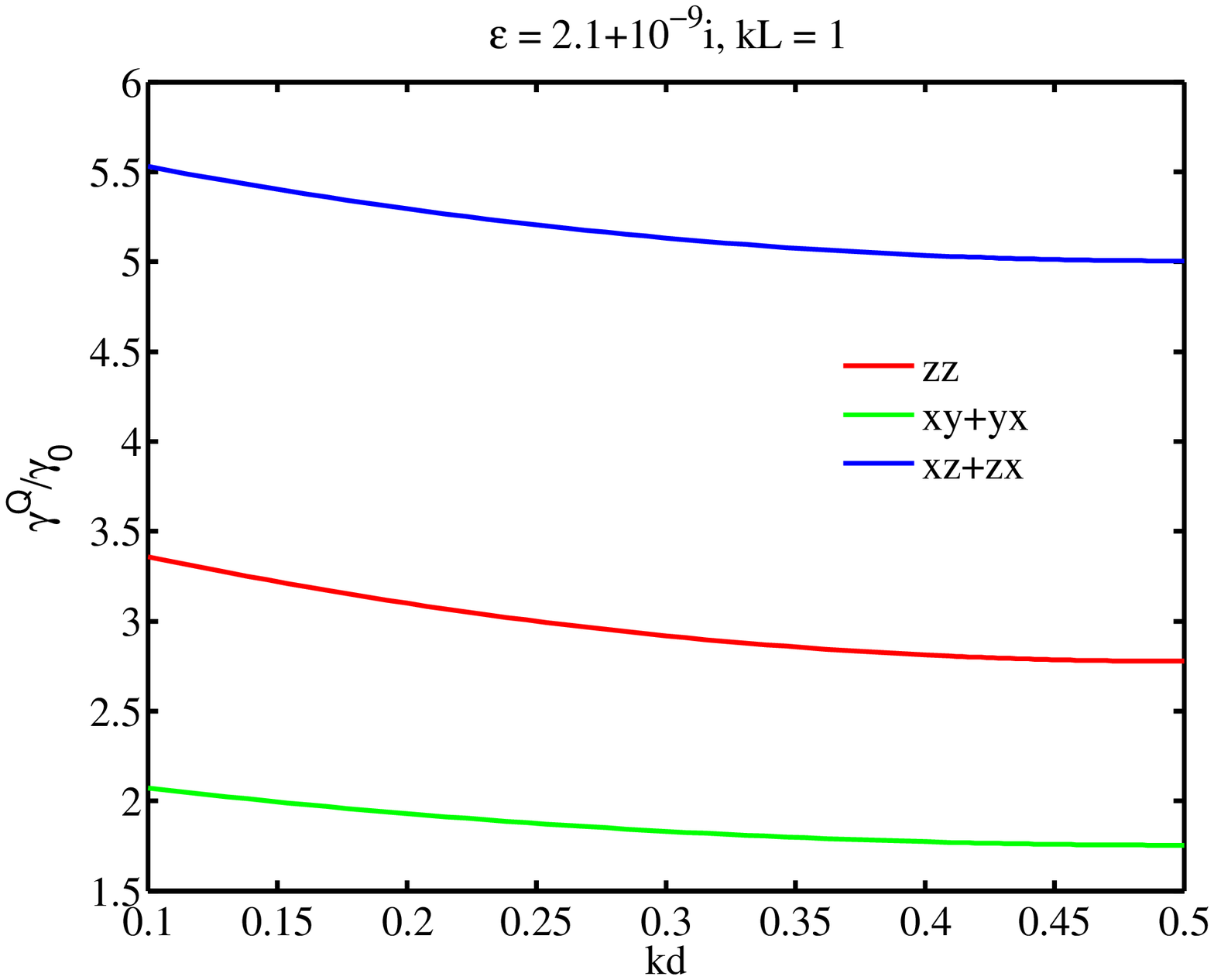}%
\caption{\label{fig7} The spontaneous decay rates of different 
quadrupoles versus their position between two quartz half-spaces with 
$\epsilon=2.1+0.000000001i $(silica) in the case of a nano-resonator ( $kL=1$). }

\end{figure}

A planar cavity can also be realized on the basis of two opposite dielectric
half-spaces. No propagating waveguiding modes are formed in that case, and
the integrand, respectively, has no poles in a complex plane near a real
axis. This should simplify a numerical calculation of the integrals. Figures
\ref{fig6},\ref{fig7}   demonstrate the dependencies of the quadrupole decay rates on the atomic
position and structure of the quadrupole moment for a planar resonator with
silica walls.

In the case of quite a large-size cavity (micro-cavity, Fig.\ref{fig6}) one can
observe an increase in the rate of spontaneous decays as an atom is
approaching the wall. In contrast to the case of metallic mirrors, such an
increase is due to the coupling of the non-propagating near fields emitted
by the quadrupole, with the propagating fields inside the dielectric
(silica). One can notice the influence of the intrinsic non-radiative
processes at a distance less than 1 nm only, because the imaginary part of
the quartz permittivity is very small at optical frequencies (see Eqs. (\ref%
{eq63})-(\ref{eq65})). At such distances one should take into account the
random inhomogeneities of the surface structure.

As the distance between dielectric walls is small (nano-cavity, Fig.\ref{fig7}), the
electric fields are near fields one, at any atom position between the walls,
and there occurs the effective field transformation into the wave
propagation over a dielectric. This provides a considerable acceleration of
the transitions. The intrinsic non-radiative decay channel is formed at
distances closer to the wall, and this is unseen on the picture.

Note that all the energy of an excited atom will be emitted in the
dielectric, and all the losses will, therefore, be non-radiative. But in the
case of the weakly absorbing dielectrics, including silica, it is not
unreasonable to distinguish between the regions of the effective
transformation into the propagating waves and the regions of the intrinsic
radiative losses.

\subsection{Atom inside ultra thin cell}

Very interesting spectroscopy experiments are carried out now with atoms
inside an extra-thin dielectric cell. Suffice it to say that the width of
the cell cab be as small as 20 nm \cite{Ducloy2}. So, it is very interesting
to understand the behavior of decay rate in that case.

All dimensional parameters are small in comparison with wavelength. As a
result we can use the quasi-static approximation to calculate decay rates
found in a previous section. The quasistatic approach here is equivalent to
the case of $k > > k_{0} = \omega / c$. In this limit the Fresnel reflection
coefficients can be simplified substantially

\begin{equation}
\begin{array}{l}
r_{12}^{p}=r_{13}^{p}=r={\dfrac{{\varepsilon -1}}{{\varepsilon +1}}} \\ 
r_{12}^{s}=r_{13}^{s}=0%
\end{array}
\label{eq59}
\end{equation}

As a result the decay rate in the small width of cavity case will have the
following form:

\begin{equation}  \label{eq60}
\left( {{\frac{{\gamma}} {{\gamma _{0}}} }} \right)_{zz}^{Q} = - {\frac{{45}%
}{{8}}}Im{\sum\limits_{n = 0}^{\infty} {r^{2n + 1}\left( {{\frac{{2r}}{{{%
\left[ {\tilde {L}\left( {n + 1} \right)} \right]}^{5}}}} - {\frac{{1}}{{{%
\left[ {\tilde {s} + \tilde {L}n} \right]}^{5}}}} - {\frac{{1}}{{{\left[ {%
\tilde {z}_{0} + \tilde {L}n} \right]}^{5}}}}} \right)}}
\end{equation}

\begin{equation}  \label{eq61}
\left( {{\frac{{\gamma}} {{\gamma _{0}}} }} \right)_{xy}^{Q} = - {\frac{{15}%
}{{16}}}Im{\sum\limits_{n = 0}^{\infty} {r^{2n + 1}\left( {{\frac{{2r}}{{{%
\left[ {\tilde {L}\left( {n + 1} \right)} \right]}^{5}}}} - {\frac{{1}}{{{%
\left[ {\tilde {s} + \tilde {L}n} \right]}^{5}}}} - {\frac{{1}}{{{\left[ {%
\tilde {z}_{0} + \tilde {L}n} \right]}^{5}}}}} \right)}}
\end{equation}

\begin{equation}  \label{eq62}
\left( {{\frac{{\gamma}} {{\gamma _{0}}} }} \right)_{xz}^{Q} = {\frac{{15}}{{%
16}}}Im{\sum\limits_{n = 0}^{\infty} {r^{2n + 1}\left( {{\frac{{2r}}{{{\left[
{\tilde {L}\left( {n + 1} \right)} \right]}^{5}}}} + {\frac{{1}}{{{\left[ {%
\tilde {s} + \tilde {L}n} \right]}^{5}}}} + {\frac{{1}}{{{\left[ {\tilde {z}%
_{0} + \tilde {L}n} \right]}^{5}}}}} \right)}}
\end{equation}

\noindent where $\tilde {z}_{_{0}} ,\tilde {s},\tilde {L}$ stand for $k_{0}
z_{0} ,k_{0} s,k_{0} L$ , respectively. In the case when atom is very close
to one surface $\tilde {z}_{0} < < \tilde {L}$ only one term ( n=0 ) is
important in this series

\begin{equation}  \label{eq63}
\left( {{\frac{{\gamma}} {{\gamma _{0}}} }} \right)_{zz}^{Q} = {\frac{{45}}{{%
8\tilde {z}_{0}^{5}}} }Imr
\end{equation}

\begin{equation}  \label{eq64}
\left( {{\frac{{\gamma}} {{\gamma _{0}}} }} \right)_{xy}^{Q} = {\frac{{15}}{{%
16\tilde {z}_{0}^{5}}} }Imr
\end{equation}

\begin{equation}  \label{eq65}
\left( {{\frac{{\gamma}} {{\gamma _{0}}} }} \right)_{xz}^{Q} = {\frac{{15}}{{%
16\tilde {z}_{0}^{5}}} }Imr
\end{equation}

From this asymptotics one can see that the total decay rate increases
inversely proportional to the fifth power of distance to surface $z=0$. This
behavior is different substantially from the dipole case, where decay rates
increase inversely proportional to the third power of distance to surface.
Another interesting point one can get from (\ref{eq63}) - (\ref{eq65}) , is
that the $zz$-quadrupoles suffer a six-fold enhancement in comparison with
other components.

\section{\textbf{Conclusions}}

In this article, the processes of the spontaneous quadrupole atomic
radiation in an arbitrary environment were considered within the framework
of both classical and quantum electrodynamics. The general equations derived
for the rates of quadrupole transitions were expressed through the spatial
derivatives of the retarded Green function corresponding to the classical
problem of electrodynamics. It was shown that the expressions differ by a
numerical coefficient 4 only, which is connected with the definition of
quadrupole moments which have different physical sense in the classical and
quantum mechanics. The expressions for the relative decay rates, i.e. the
rates normalized by the uniform space rate, prove to be identical.

The results obtained are applied to a description of quadrupole atomic
transitions in a planar cavity. The explicit analytical expressions for the
rates of any quadrupole transition were found for such a cavity. The results
have been analyzed in detail for the planar cavities with dielectric and
metallic walls. It was found that the quadrupole transitions are accelerated
with decreasing resonator size. In the case of dielectric walls, such an
acceleration is due to the transformation of the near dipole fields into the
propagating waves inside the dielectric. In the case of metallic mirrors,
the acceleration becomes more profound, and is due to the radiation
absorption at the surface layer of a metal.

In this paper we restrict ourselves to investigation of quadrupole decay
rates. However, our approach can be also applied to description of frequency
shifts of quadrupole transitions in nanoenviroment. Again, general
expressions for frequency shift will be expressed through space derivatives
of retarded Green function. We will present detailed investigation of
frequency shifts of quadrupole transitions in nanoenviroment in a separate
publication.

\begin{acknowledgments}

The authors thank the Russian Foundation for Basic Research, grant \#
04-02-16211 (V.K.), and Centre National de la Recherche Scientifique (V.K.,
M.D.) for their financial support of this work. One of the authors (V.K.) is
grateful to the colleagues of the Laboratoire de Physique des Lasers
(Universite Paris-Nord), where this work has been completed, for their
hospitality. This work has been done as part of the European Union FASTNET
consortium.

\end{acknowledgments}

\newpage

\newpage

\newpage

\newpage

\newpage

\newpage

\newpage

\end{document}